\documentclass[onecolumn,showpacs]{revtex4}

\usepackage{graphicx}% Include figure files
\usepackage{dcolumn}% Align table columns on decimal point
\usepackage{bm}% bold math
\usepackage{amssymb}
\usepackage{amsmath}
\usepackage[colorlinks=true, urlcolor=blue,linkcolor=blue,citecolor=blue]{hyperref}

\input epsf

\begin{document}

\title{\Large  Variable Modified Chaplygin Gas in Anisotropic Universe with Kaluza-Klein Metric}

\author{\bf Chayan Ranjit$^1$\footnote{chayanranjit@gmail.com} Shuvendu Chakraborty$^2$\footnote{shuvendu.chakraborty@gmail.com} and Ujjal
Debnath$^3$\footnote{ujjaldebnath@yahoo.com,
ujjal@iucaa.ernet.in}}

\affiliation{$^{1,2}$Department of Mathematics, Seacom Engineering College, Howrah - 711 302, India.\\
$^3$Department of Mathematics, Bengal Engineering and Science
University, Shibpur, Howrah-711 103, India.}

\date{\today}

\begin{abstract}
In this work, we have consider Kaluza-Klein Cosmology for
anisotropic universe where the universe is filled with variable
modified chaplygin gas (VMCG). Here we find normal scalar field
$\phi$ and the self interacting potential $V(\phi)$ to describe
the VMCG Cosmology. Also we graphically analyzed the geometrical
parameters named {\it statefinder parameters} in anisotropic
Kaluza-Klein model. Next, we consider a Kaluza-Klein model of
interacting VMCG with dark matter in the Einstein gravity
framework. Here we construct the three dimensional autonomous
dynamical system of equations for this interacting model with the
assumption that the dark energy and the dark matter are interact
between them and for that we also choose the interaction term. We
convert that interaction terms to its dimensionless form and
perform stability analysis and solve them numerically. We obtain a
stable scaling solution of the equations in Kaluza-Klein model and
graphically represent solutions.
\end{abstract}


\maketitle

\section{\normalsize\bf{Introduction}}
The recent observations of the luminosity of type Ia supernovae
\cite{Bachall et al. 1999, Perlmutter et al. 1999} indicate that
the universe is currently undergoing an accelerated expansion. The
theory of Dark energy having negative pressure is the main
responsible for this scenario. From recent cosmological
observations including supernova data \cite{A.G.Riess et al. 2007}
and measurements of cosmic microwave background radiation (CMBR)
\cite{D.N.Spergel et al} it is evident that our present universe
is made up of about 2/3 dark energy and about 1/3 dark matter of
the whole universe. There are several interesting mechanisms such
as higher dimensional phenomena \cite{Debnath2010}, Loop Quantum
Cosmology (LQC) \cite{Eisenstein 2005}, modified gravity
\cite{Cognola 2009}, Brans-Dicke theory \cite{Brans 1961},
brane-world model \cite{Maartens 2001} and many others which are
extremely applicable in recent research to explain the present
scenario of the universe.\\

A fundamental theory on higher dimensional model was introduced by
Kaluza-Klein \cite{T. Kaluza,O. Klein}, where they considered an
extra dimension with FRW metric to unify Maxwell's theory of
electromagnetism and Einstein's gravity. After extensive research,
scientists are realized that higher dimensional Cosmology may be
more useful to understand the interaction of particles. Also to
study the universe, as our space-time is explicitly four
dimensional in nature, the `hidden' dimensions must be related to
the dark matter and dark energy. In modern physics Kaluza-Klein
theory showed its great importance in string theory
\cite{Polchinski 1998}, in supergravity \cite{Duff et al. 1986}and
in superstring theories \cite{Green et al. 1987}. Li et al
\cite{Li et al. 1998} have considered the inflation in
Kaluza-Klein theory. In 1997, Overduin and Wesson \cite{Overduin
et al. 1997} represented a review of Kaluza-Klein theory with
higher dimensional unified theories. String cloud and domain walls
with quark matter in N-dimensional Kaluza-Klein Cosmological model
was presented by Adhav et al \cite{Adhav 2008}. Recently some
authors \cite{Qiang 2005,Chen 2009,Ponce 1988,Chi 1990,Fukui
1993,Liu 1994,Coley 1994} studied Kaluza-Klein cosmological models
with different dark energies and dark matters.\\

Recently dynamical system theory become a very important tool to
study cosmology and astrophysics in the framework of general
relativity. Some authors \cite{Santos 1997,Romero,Kolitch
1996,Holden 1998,Kolitch 1995} work on qualitative analysis in
dynamical system of FRW Cosmology in Brans-Dicke gravity, where
they showed that the field equations of FRW may be reduced to two
dimensional dynamical system whether in presence of scalar
potential it could be reduced in three dimensional.\\

In this work, we have considered Kaluza-Klein Cosmology for
Anisotropic Universe where the Universe is filled with variable
modified Chaplygin gas (VMCG). Here we find normal scalar field
$\phi$ and the self interacting potential $V(\phi)$ to describe
the VMCG Cosmology. Also we graphically analyzed the geometrical
parameters named {\it statefinder parameters} in anisotropic
Kaluza-Klein model. In the later section, we consider a
Kaluza-Klein model of interacting VMCG with dark matter in the
Einstein gravity framework. Here we construct the three
dimensional autonomous dynamical system of equations for this
interacting model with the assumption that the dark energy and the
dark matter are interact between them and for that we also choose
the interaction term. We convert that interaction terms to its
dimensionless form and perform stability analysis and solve them
numerically. We obtain a stable scaling solution of the equations
in Kaluza-Klein model and graphically represent solutions.\\

\section{\normalsize\bf{Basic Equations and Solutions}}

The Kaluza-Klein type metric is given by \cite{Adhav 2011}
\begin{equation}
ds^{2}=dt^{2}-a^{2}(dx^{2}+dy^{2}+dz^{2})-b^{2}d\psi^{2}
\end{equation}
where $a$ and $b$ are function of cosmic time $t$ only. Here we
are dealing only with an anisotropic fluid whose energy momentum
tensor is in the following form:
\begin{center}
$T_{v}^{u}=diag[T_{0}^{0},T_{1}^{1},T_{3}^{3},T_{4}^{4}]$
\end{center}
We parameterize it as follows:
\begin{eqnarray*}
T_{v}^{u}=diag[\rho,-p_{x},-p_{y},-p_{z},-p_{\psi}]
\end{eqnarray*}
\begin{eqnarray*}
\qquad\:=diag[1,-\omega_{x},-\omega_{y},-\omega_{z},-\omega_{\psi}]\rho
\end{eqnarray*}
\begin{equation}
\:=diag[1,-\omega,-\omega,-\omega,-\omega]\rho
\end{equation}
where $\rho$ is the energy density of the fluid; $p_{x}, p_{y},
p_{z}$ and $p_{\psi}$ are the pressures and $\omega_{x},
\omega_{y}, \omega_{z}$ and $\omega_{\psi}$ are the directional
equation of state $(EoS)$ parameters of the fluid.\\

The Einstein field equations, in natural limits $(8\pi G=c=1)$ are
\begin{equation}
G_{\mu \nu}=R_{\mu \nu}-\frac{1}{2}R g_{\mu \nu}=-T_{\mu \nu}
\end{equation}
where $g_{\mu \nu}u^{\mu}u^{\nu}=1;u^{\mu}=(1,0, 0, 0, 0)$ is the
velocity vector; $R_{\mu \nu}$ is the Ricci tensor; R is the Ricci
scalar, $T_{\mu \nu}$ is the energy momentum tensor. In a
co-moving coordinate system, Einstein's filled equation (3), for
the anisotropic Kaluza-Klein spacetime (1), with eq.(2) (for flat
universe i.e., for $k=0$) yield
\begin{equation}
3\frac{\dot{a}\dot{b}}{ab}+3\frac{\dot{a}^{2}}{a^{2}}=\rho
\end{equation}
\begin{equation}
2\frac{\ddot{a}}{a}+\frac{\ddot{b}}{b}+\frac{\dot{a}^{2}}{a^{2}}
+2\frac{\dot{a}\dot{b}}{ab}=-p
\end{equation}
\begin{equation}
3\frac{\ddot{a}}{a}+3\frac{\dot{a}^{2}}{a^{2}}=-p
\end{equation}
where the overhead dot$(\,\dot{ }\,)$ denote derivative with
respect to the cosmic time t. The energy conservation equation in
Kaluza-Klein Cosmological Model is as
 \begin{equation}
 \dot{\rho}+4H(\rho+p)=0
 \end{equation}

In the case of {\it pure Chaplygin gas} obeys an equation of state
like \cite{Kamenshchik et al. 2001,Gorini et al. 2004a}
\begin{equation}
p=-\frac{B}{\rho}
\end{equation}
where $p$ and $\rho$ are respectively the pressure and energy
density and $B$ is a positive constant. Subsequently in the case
of {\it generalized Chaplygin gas} \cite{Gorini et al. 2003,Alam
et al. 2003,Bento et al. 2002} obeys an equation of state
\begin{equation}
p=-\frac{B}{\rho^{\alpha}}\quad\quad with\quad 0\leq \alpha \leq 1
\end{equation}
 and in the case of {\it modified Chaplygin gas}
 \cite{Benaoum 2002,Debnath et al 2004} obeys an equation of state
\begin{equation}
p=A\rho-\frac{B}{\rho^{\alpha}}\quad\quad with\quad 0\leq
\alpha\leq 1
\end{equation}
where $A$, $B$ are positive constants. This equation of state
shows that radiation era (when $A=1/3$) at one extreme (when the
scale factor $a(t)$ is vanishingly small) while a $\Lambda CDM$
model at other extreme (when the scale factor $a(t)$ is infinitely
large).

{\it Guo and Zhang} $2005a$ \cite{Guo and Zhang 2005a} first
proposed {\it variable Chaplygin gas} model with equation of state
$(8)$, where $B$ is a positive function of the cosmological scale
factor $`a$' i.e., $B=B(a)$. This assumption is reasonable since
$B(a)$ is related to the scalar potential if we take the Chaplygin
gas as a Born- Infeld scalar field {\it Bento et al. 2003}
\cite{Bento et al. 2003}. There are some works on variable
Chaplygin gas model like as {\it Sethi et al.} $2005$ \cite{Sethi
et al 2005}, {\it Guo and Zhang} $2005b$ \cite{Guo and Zhang
2005b}. In 2007, {\it Debnath } \cite{Debnath 2007} introduced
VMCG with equation of state $(10)$ where B is a positive function
of the cosmological scale factor $`a$' $(i.e., B = B(a))$ as
\begin{equation}
p=A\rho-\frac{B(a)}{\rho^{\alpha}}\quad\quad with\quad 0\leq
\alpha\leq 1
\end{equation}
where $A$ is a positive constant. Now, we have considered that the
anisotropic universe filled with VMCG obeys equation of state (10)
where $B$ is a positive function of the cosmological scale factors
$a$ and $b$ (i.e., $B=B(a,b)$)

\begin{equation}
p=A\rho-\frac{B(a,b)}{\rho^{\alpha}}\quad\quad with\quad 0\leq
\alpha\leq 1
\end{equation}
Now, for simplicity, assume $B(a,b)$ is of the form
\begin{equation}
B(a,b)=B_{0}\left(a^{3}b\right)^{-n}
\end{equation}
where $B_{0}> 0$ and $n$ are constants. From (7) using $(12)$ and
$(13)$ we have

\begin{equation}
\rho=\left[\frac{B_{0}(\alpha +1)}{\left((A+1)(\alpha
+1)-n\right)}\frac{1}{\left(a^{3}b\right)^{n}}+\frac{C}{(a^{3}b)^{(\alpha
+1)(A+1)}}\right]^{\frac{1}{\alpha +1}}
\end{equation}
where $C > 0$ is an arbitrary integration constant and
$(1+A)(1+\alpha)> n$, for positivity of first term. Here $n$ must
be positive, because otherwise, $a\rightarrow \infty$ implies
$\rho \rightarrow \infty$, which is not the case for expanding
Universe. Now, we consider the energy density and pressure
corresponding to normal scalar field $\phi$ having a
self-interacting potential $V (\phi)$ as follows:
\begin{equation}
\rho_{\phi}=\frac{1}{2}\dot{\phi}^{2}+V(\phi)=\rho=
\left[\frac{B_{0}(\alpha +1)}{\left((A+1)(\alpha
+1)-n\right)}\frac{1}{\left(a^{3}b\right)^{n}}+\frac{C}{(a^{3}b)^{(\alpha
+1)(A+1)}}\right]^{\frac{1}{\alpha +1}}
\end{equation}
and
\begin{eqnarray*}
p_{\phi}=\frac{1}{2}\dot{\phi}^{2}-V(\phi)=A\rho-
\frac{B_{0}(a^{3}b)^{-n}}{\rho^{\alpha}}=
A\left[\frac{B_{0}(\alpha+1)}{\left((A+1)
(\alpha+1)-n\right)}\frac{1}{\left(a^{3}b\right)^{n}}
+\frac{C}{(a^{3}b)^{(\alpha+1)(A+1) }}\right]^{\frac{1}{\alpha+1}}
\end{eqnarray*}
\begin{equation}
-\frac{B_{0}(a^{3}b)^{-n}}{\left[\frac{B_{0}(\alpha+1)
}{\left((A+1)(\alpha+1)-n\right)}\frac{1}{\left(a^{3}b\right)^{n}}+
\frac{C}{(a^{3}b)^{(\alpha+1)(A+1)
}}\right]^{\frac{\alpha}{\alpha+1}}}
\end{equation}
Hence for flat universe $(i.e., k = 0)$, we have
\begin{eqnarray*}
\phi=\int{\left[(A+1)\left(\frac{B_{0}(\alpha+1)(a^{3}b)^{-n}}{1+A-n+\alpha+A
\alpha}+ C
\left(a^{3}b\right)^{-(A+1)(\alpha+1)}\right)^{\frac{1}{\alpha+1}}
\right.}
\end{eqnarray*}
\begin{equation}
\left.-B_{0}(a^{3}b)^{-n}\left(\left(\frac{B_{0}(\alpha+1)(a^{3}b)^{-n}}{1+A-n+\alpha+A
\alpha}+ C
\left(a^{3}b\right)^{-(A+1)(\alpha+1)}\right)^{\frac{1}{\alpha+1}}
\right)^{-\alpha}\right]^{\frac{1}{2}}~dt
\end{equation}
 and
\begin{eqnarray*}
V=\frac{1}{2}{\left[-(A-1)\left(\frac{B_{0}(\alpha+1)
(a^{3}b)^{-n}}{1+A-n+\alpha+A \alpha}+ C
\left(a^{3}b\right)^{-(A+1)(\alpha+1)}\right)^{\frac{1}{\alpha+1}}
\right.}
\end{eqnarray*}
\begin{equation}
\left.-B_{0}(a^{3}b)^{-n}\left(\left(\frac{B_{0}(\alpha+1)(a^{3}b)^{-n}}{1+A-n+\alpha+A
\alpha}+ C
\left(a^{3}b\right)^{-(A+1)(\alpha+1)}\right)^{\frac{1}{\alpha+1}}
\right)^{-\alpha}\right]
\end{equation}
\begin{figure}
\includegraphics[scale=0.5]{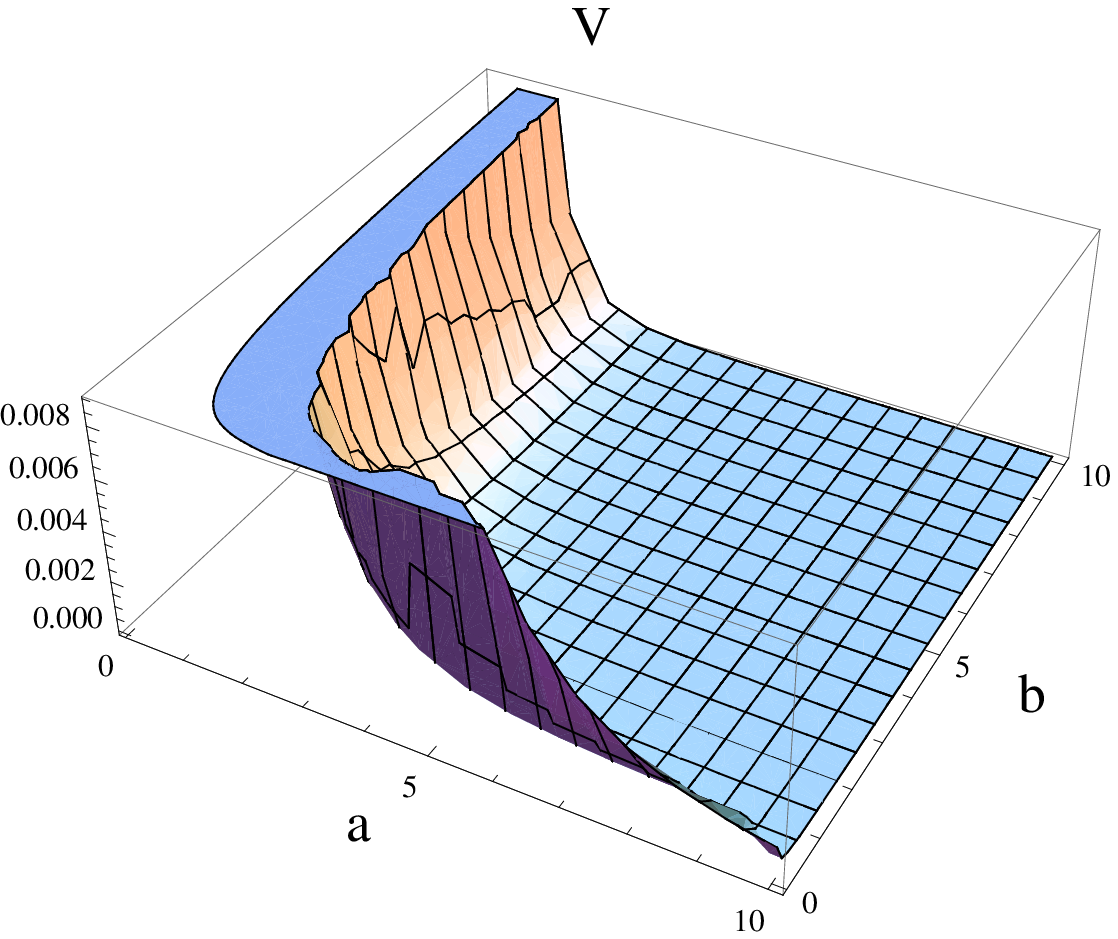}~~~~~~~~~~~~~~~
\includegraphics[scale=0.5]{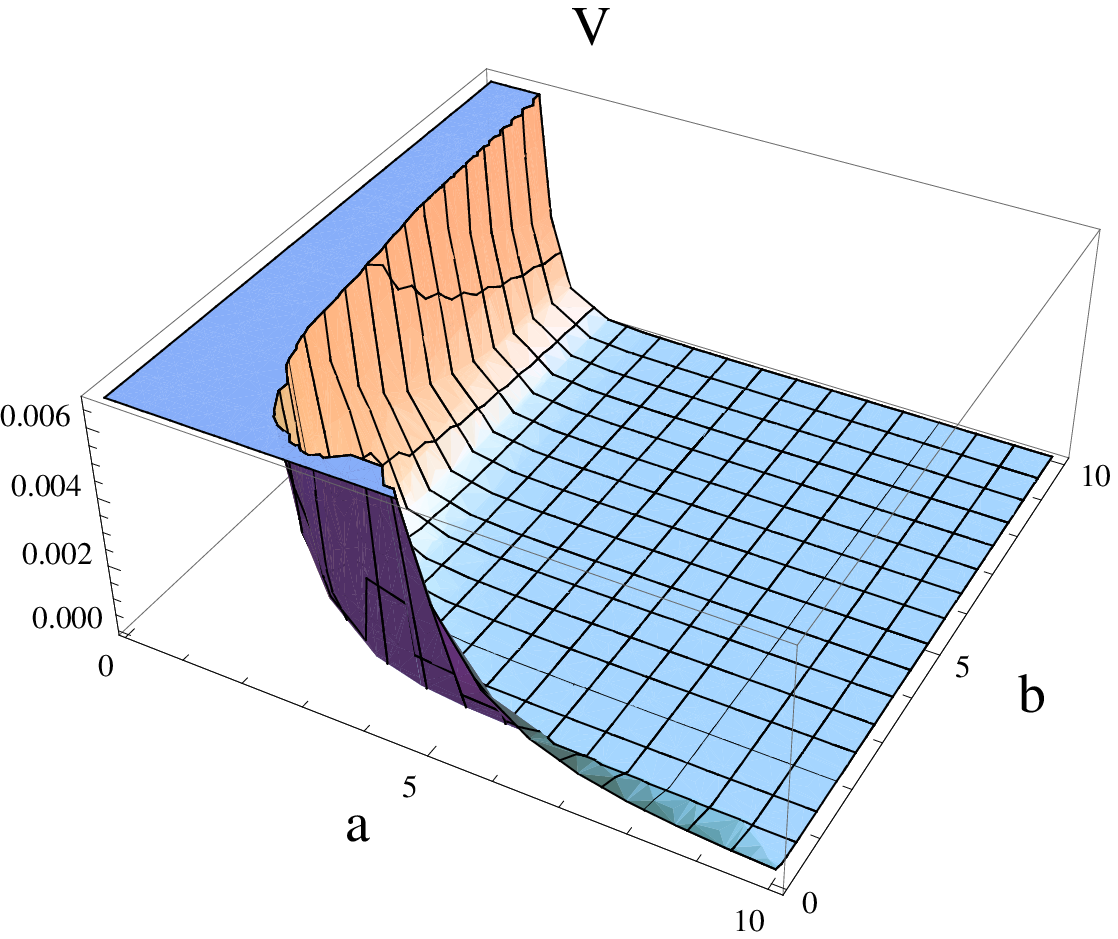}\\
\vspace{2mm}
~~~~~~~~Fig.1~~~~~~~~~~~~~~~~~~~~~~~~~~~~~~~~~~~~~~~~~~~~~~~~~~~~~~~~~~~~~~~Fig.2\\
\vspace{4mm}
\includegraphics[scale=.5]{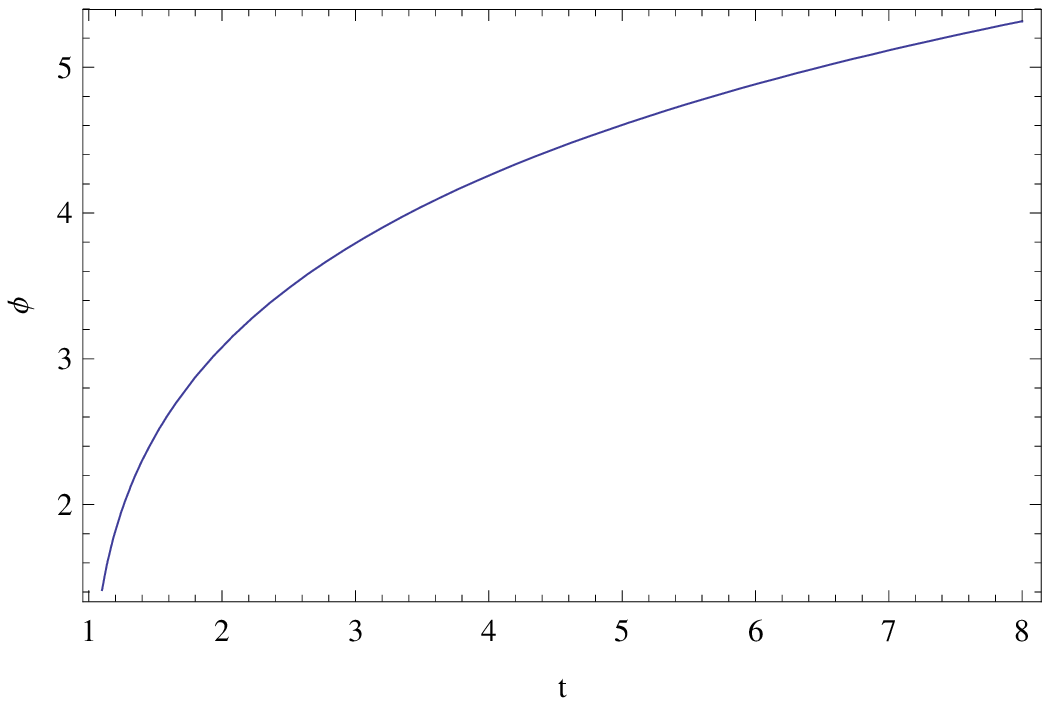}\\
\vspace{2mm}
~~~~~~~~~~~~~~~~~~~~~~~~~~~~~~~~~~~~Fig.3~~~~~~~~~~~~~~~~~~~~~~~~~~~~~~~~\\
\vspace{6mm}

Fig. 1 and Fig. 2 show the variations of $V$ against $a$ and $b$
for $A=\frac{1}{3}$ and $A=1$ respectively with $B=1, C=2, n=3,
\alpha=0.8$ and Fig. 3 shows the variation of $\phi$ against $t$
for $A=\frac{1}{3}, B=2, C=1, n=1.1, d=1, \alpha=0.5$ in the case
of the anisotropic universe filled with {\it Variable Modified
Chaplygin Gas} with  Kaluza-Klein metric. \vspace{6mm}
\end{figure}

From equation (17), we see that the integration is very
complicated. So the numerical investigations are needed to get the
physical interpretation of the scalar field and the corresponding
potential. Fig. 1 and Fig. 2 show the variations of $V$ against
$a$ and $b$ for $A=\frac{1}{3}$ and $A=1$ respectively with $B=1,
C=2, n=3, \alpha=0.8$ and Fig. 3 shows the variation of $\phi$
against $t$ for $A=\frac{1}{3}, B=2, C=1, n=1.1, d=1, \alpha=0.5$
in the case of the anisotropic universe filled with {\it Variable
Modified Chaplygin Gas} with Kaluza-Klein metric. From the
figures, we see that $\phi$ increases due to evolution of the
universe. Also the potential $V$ decreases with the scale factors
$a$ and $b$.\\

In the paper Gorini et al. 2003 \cite{Gorini et al. 2003}; Alam et
al. 2003 \cite{Alam et al. 2003}, the flat Friedmann model filled
with Chaplygin fluid has been analyzed in terms of the recently
proposed "statefinder" parameters Sahni et al. 2003 \cite{Sahni et
al. 2003}. The statefinder diagnostic along with future SNAP
observations may perhaps be used to discriminate between different
dark energy models. The above statefinder diagnostic pair for
higher dimensional anisotropic cosmology are constructed from the
scale factors $a$ and $b$ as follows:

\begin{equation}
r=1+3\frac{\dot{H}}{H^{2}}+\frac{\ddot{H}}{H^{3}} ~~\text{and} ~~
s=\frac{r-1}{3(q-\frac{1}{2})}
\end{equation}
where $q$ is the deceleration parameter defined by
$q=-1-\frac{\dot{H}}{H^{2}}$ and $H$ is the Hubble parameter.
Since this parameters are dimensionless so they allow us to
characterize the properties of dark energy in a model
independently. Finally we graphically analyzed geometrical
parameters $r$ and $s$ with respect to $a$ and $b$ in the case of
the anisotropic universe filled with {\it Variable Modified
Chaplygin Gas}.

For Kaluza-Klein model with flat universe $(i.e., k = 0)$ we have
\begin{equation}
4 \dot{H}+16 H^{2}=\frac{4}{3}(\rho-p)
\end{equation}
Therefore
\begin{equation}
q=3-\frac{\rho-p}{3H^{2}}
\end{equation}
Hence
\begin{equation}
r=21+\frac{4(\rho-p)}{3H^{2}}\left[\frac{7}{4}+(\alpha+1)A-\frac{\alpha
p}{\rho}\right]+\frac{4n}{3H^{2}}(p-A\rho)
\end{equation}
and
\begin{equation}
s=\frac{4}{3(\frac{15}{2}-\frac{p-\rho}{H^{2}})}\left[15+\frac{\rho-p}{H^{2}}
(\frac{7}{4}+(\alpha+1)A-\frac{\alpha
p}{\rho})+\frac{n}{H^{2}}(p-A\rho)\right]
\end{equation}
where $\rho$  and $p$ are given by $(15)$ and $(16)$. Using field
equations, $r$ and $s$ can be expressed in terms of scale factors
$a$ and $b$ in very complicated form, so we have not presented the
explicit expressions here. Fig. 4 and Fig. 5 show the variations
of $r$ against $a$ and $b$ and the variations of $s$ against $a$
and $b$ and Fig. 6 shows the variation of $r$ against $s$ for
$A=1, B=1, C=2, n=3, \alpha=0.8$ and $H=1$ respectively in the
case of the anisotropic universe filled with {\it Variable
Modified Chaplygin Gas}. From the figures, we conclude that $r$
increases and $s$ decreases as $a$ and $b$ increase. From figure
6, we see that $s$ increases from some positive value to $+\infty$
as $r$ decreases from finite positive range. Also after certain
stage, $s$ increases from $-\infty$ to $0$ as $r$ decreases from
some positive to negative value.\\

\begin{figure}
\includegraphics[scale=0.5]{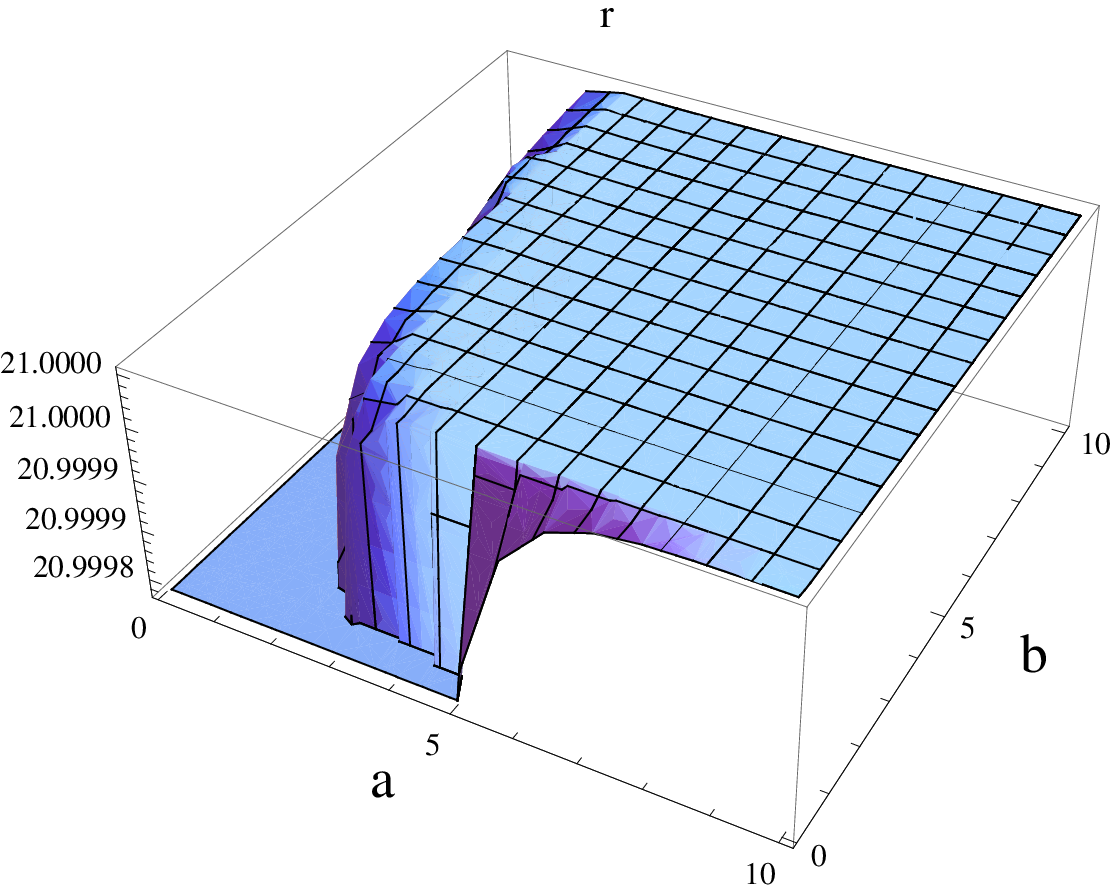}~~~~~~~~~~~~~~~
\includegraphics[scale=0.55]{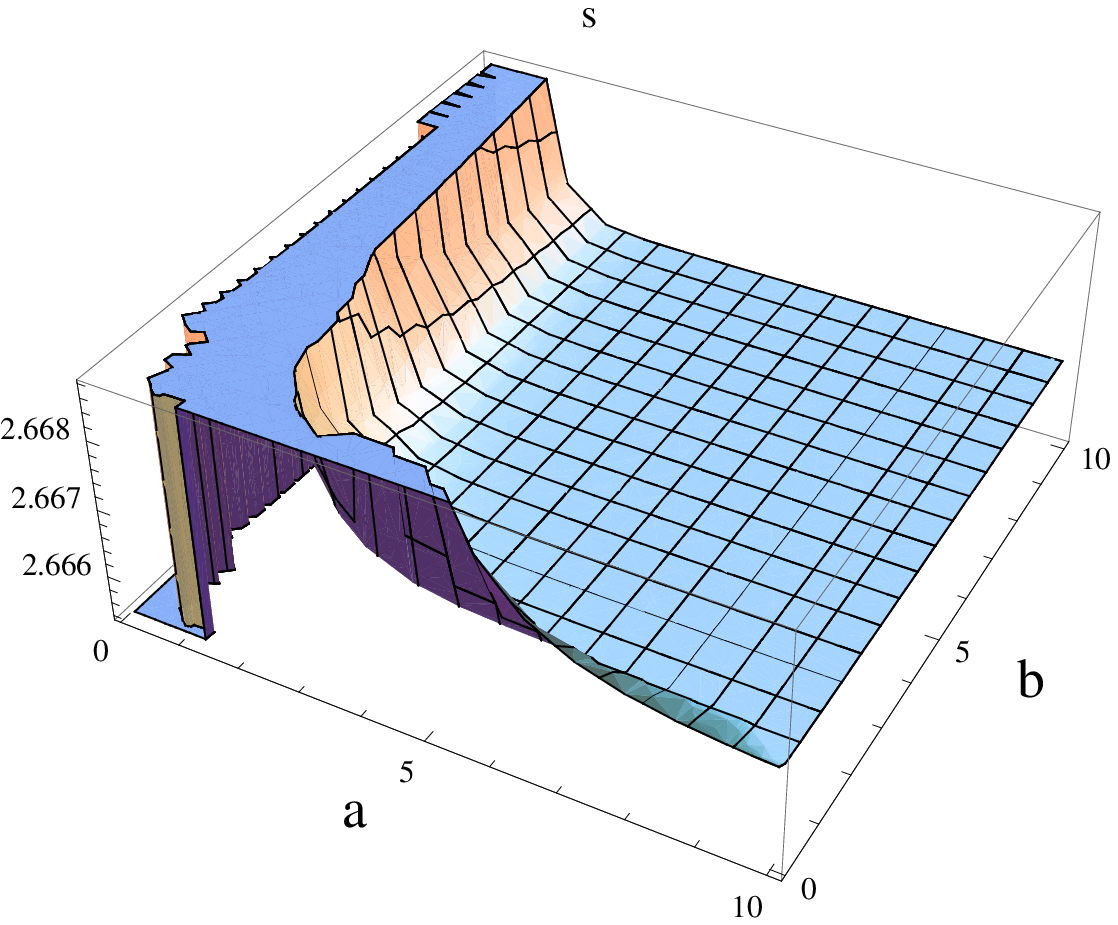}\\
\vspace{2mm}
~~~~~~~~Fig.4~~~~~~~~~~~~~~~~~~~~~~~~~~~~~~~~~~~~~~~~~~~~~~~~~~~~~~~~~~~~~~~Fig.5\\
\vspace{4mm}
\includegraphics[scale=.55]{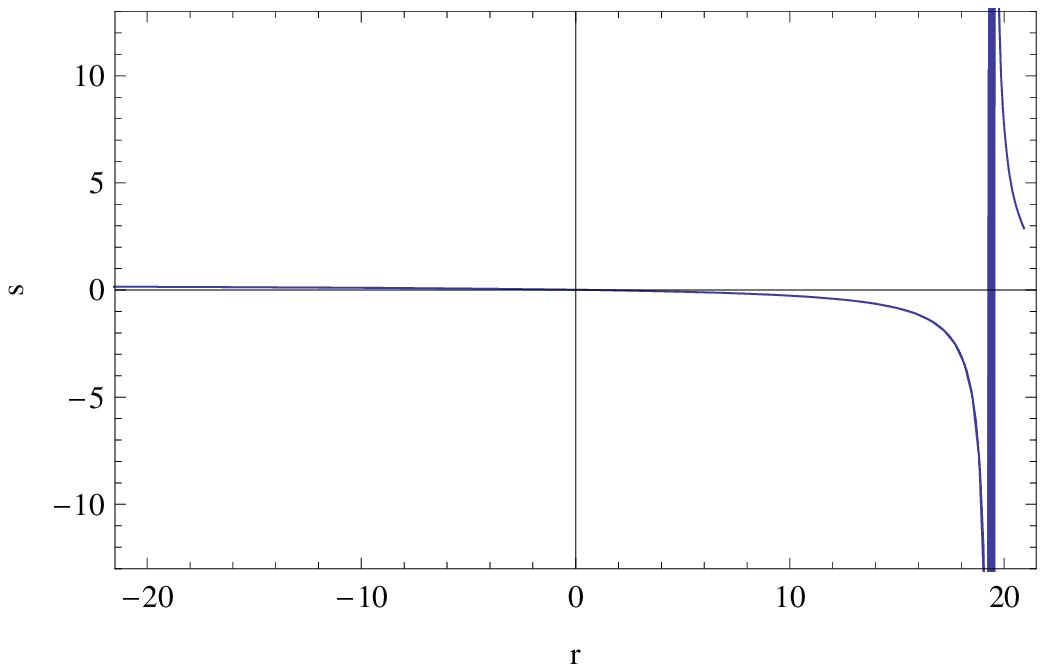}\\
\vspace{2mm}
~~~~~~~~~~~~~~~~~~~~~~~~~~~~~~~~~~~~Fig.6~~~~~~~~~~~~~~~~~~~~~~~~~~~~~~~~\\
\vspace{6mm} Fig. 4 and Fig. 5 show the variations of $r$ against
$a$ and $b$ and the variations of $s$ against $a$ and $b$ for
$A=1, B=1, C=2, n=3, \alpha=0.8$ and $H=1$ respectively and Fig. 6
shows the variation of $r$ against $s$ for $A=1, B=1, C=2, n=3,
\alpha=0.8$ and $H=1$ respectively in the case of the anisotropic
universe filled with {\it Variable Modified Chaplygin Gas}.
\vspace{6mm}
\end{figure}

\section{\normalsize\bf{DYNAMICAL SYSTEM ANALYSIS IN KALUZA-KLEIN MODEL WITH INTERACTING VMCG AND DARK MATTER}}

We consider the universe filled with VMCG as dark energy and dark
matter interacting through an interaction term. Due to interaction
between the two components, the energy conservation would not hold
for the individual components, therefore the above conservation
equation $(7)$ will break into two non-conserving equations:
\begin{equation}
\dot{\rho}_{vmcg}+4H(\rho_{vmcg}+p_{vmcg})=-Q
\end{equation}
and
\begin{equation}
\dot{\rho}_{dm}+4H(\rho_{dm}+p_{dm})=Q
\end{equation}\\

Here $Q$ is the interaction term which have dimension of density
multiplied by Hubble parameter $H$. For this scenario the suitable
choice of $Q$ be $Q = 4bH\rho$, $b$ is the coupling parameter
denoting the transfer strength, $\rho=\rho_{vmcg}+\rho_{dm}$ and
$p=p_{vmcg}+p_{dm}$ ($p_{dm}$ is very small quantity, somewhere
dark matter assumed as pressureless quantity) are the total cosmic
energy density and pressure respectively which satisfy the above
conservation equation (7). The subscripts $vmcg$ and $dm$ denote
the VMCG and dark matter respectively.\\

The dark matter equation of state is assumed to be linear EoS
\begin{equation}
p_{dm}=\omega_{dm}\rho_{dm}
\end{equation}\\

To analyze the dynamical system, we convert the physical parameter
into dimensionless form as follows:

\begin{equation}
x=\ln(a^{3}b),~~~~ u=\frac{\rho_{vmcg}}{4H^{2}},~~~~
v=\frac{p_{vmcg}}{4H^{2}},~~~~ w=\frac{\rho_{dm}}{4H^{2}}
\end{equation}\\

The equation of state of the VMCG can be expressed as
\begin{equation}
\omega_{vmcg}=\frac{p_{vmcg}}{\rho_{vmcg}}=\frac{v}{u}
\end{equation}\\

Now we can cast the evolution equations in the following
autonomous system of $u$, $v$ and $w$ in the form:

\begin{equation}
\frac{du}{dx}=u-v-cu-cw-\frac{2}{3}u(u-v+(1-\omega_{dm})w)
\end{equation}
\begin{equation}
\frac{dv}{dx}=((1+\alpha)A-\alpha\frac{v}{u})(-u-v-c(u+w))+n(Au-v)-
\frac{2}{3}v(u-v+(1-\omega_{dm})w)+2v
\end{equation}
\begin{equation}
\frac{dw}{dx}=-(1+\omega_{dm})w+c(u+w)-\frac{2}{3}w(u-v+(1-\omega_{dm})w)+2w
\end{equation}

For the mathematical simplicity, we work out $\alpha=1$ and
$A=\frac{1}{3}$ only. The critical points of the above system are
obtained by putting $\frac{du}{dx}=\frac{dv}{dx}=\frac{dw}{dx}=0$
which yield

\begin{equation}
u_{crit}=\frac{3}{2}+\frac{3c}{n-2(1+\omega_{dm})}
\end{equation}
\begin{equation}
v_{crit}=\frac{3(n^{2}-2n(2+\omega_{dm})+4(1+\omega_{dm}+c\omega_{dm}))}{4(n-2(1+\omega_{dm}))}
\end{equation}
\begin{equation}
w_{crit}=\frac{3c}{2-n+2\omega_{dm}}
\end{equation}

For the stability of the dynamical system about the critical
point, we linearize the governing equation around the critical
point and arrived at

\begin{equation}
\delta\left(\frac{du}{dx}\right)=\left[\frac{1}{3}
\left(3-3c-4u+2v+2w(-1+\omega_{dm})\right)\right]_{crit}\delta u +
\left[\frac{2u}{3}-1\right]_{crit} \delta v+
\left[\frac{1}{3}\left(-3c+2u(-1+\omega_{dm})\right)\right]_{crit}
\delta w
\end{equation}

\begin{eqnarray*}
\delta\left(\frac{dv}{dx}\right)=\left[-\frac{3Au^{2}(1+c-n+\alpha+c\alpha)+
v(2u^{2}+3(v+cw)\alpha)}{3u^{2}}\right]_{crit}\delta u +
\left[2-n-\frac{2u}{3}+\frac{4v}{3}+(1+c)\alpha\right.
\end{eqnarray*}
\begin{equation}
\left.+\frac{(2v+cw)\alpha}{u}-A(1+\alpha)+\frac{2}{3}w(-1+\omega_{dm})
\right]_{crit} \delta v+
\left[-Ac(1+\alpha)+\frac{1}{3}v(-2+\frac{3c\alpha}{u}+2\omega_{dm})\right]_{crit}
\delta w
\end{equation}

\begin{equation}
\delta\left(\frac{dw}{dx}\right)=\left[c-\frac{2w}{3}\right]_{crit}\delta
u+\left[\frac{2w}{3}\right]_{crit}\delta v +
\left[\frac{1}{3}(3c-2u+2v+(-3+4w)(-1+\omega_{dm}))\right]\delta w
\end{equation}

The corresponding eigenvalues of the coefficient matrix of the
above equations are

\begin{equation}
\lambda_{1}=\frac{n-4}{2}
\end{equation}
and
\begin{eqnarray*}
\lambda_{2,3}=\frac{44-40n+9n^{2}+4c(3n-8)+56\omega_{dm}-
24n\omega_{dm}+12\omega_{dm}^{2}}{12(2c+n-2(1+\omega_{dm}))}
\end{eqnarray*}
\begin{equation}
\pm\frac{\sqrt{(8c(3\omega_{dm}-1)+(6\omega_{dm}+3n-10)(n-
2(\omega_{dm}+1)))(-48c^{2}+8c(9\omega_{dm}-3n+5)+
(6\omega_{dm}+3n-10)(n-2(\omega_{dm}+1)))}}{12(2c+n-2(1+\omega_{dm}))}
\end{equation}

\begin{figure}
\includegraphics[scale=0.35]{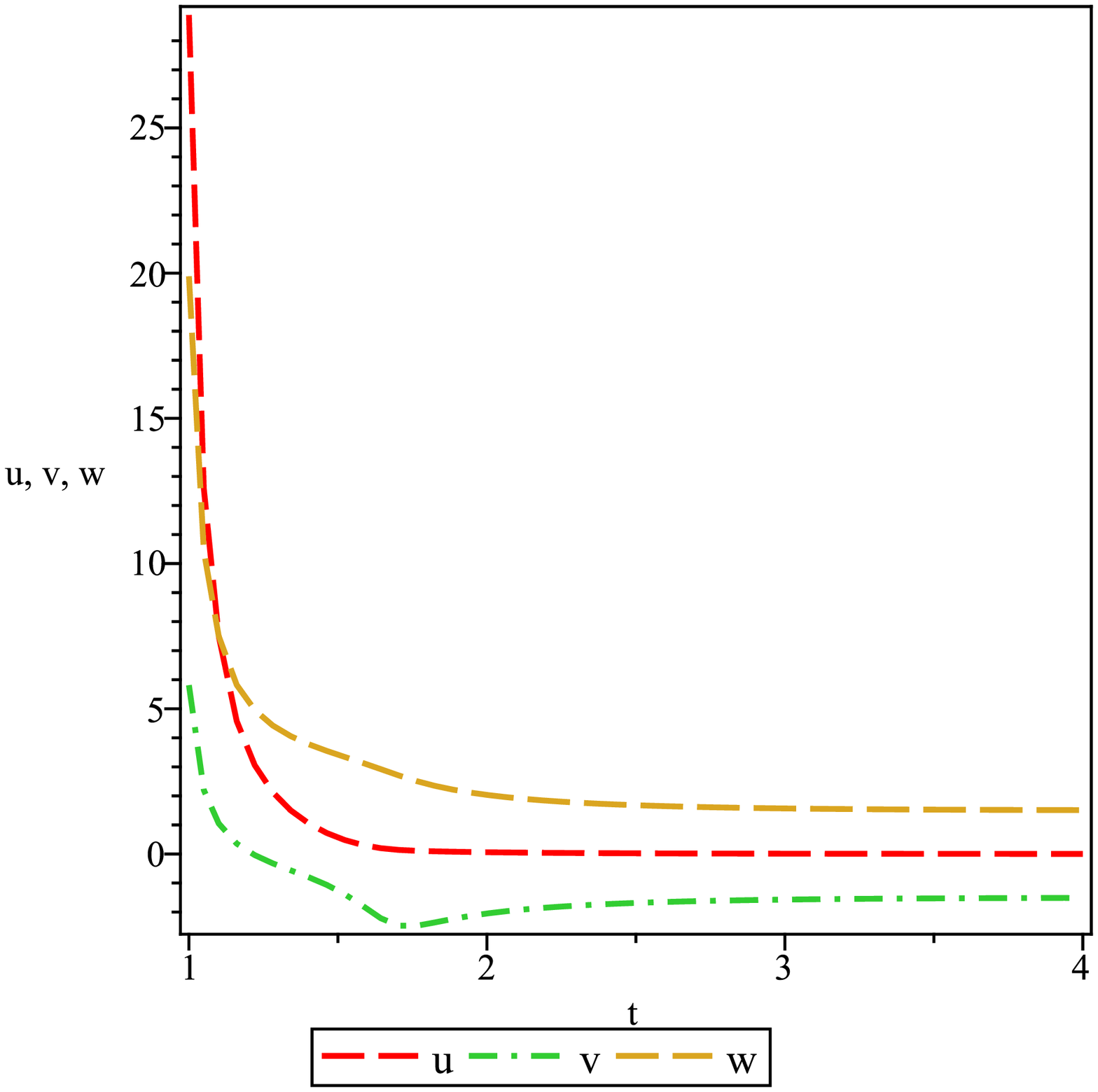}~~~~~~~~~~~~~~~
\includegraphics[scale=0.35]{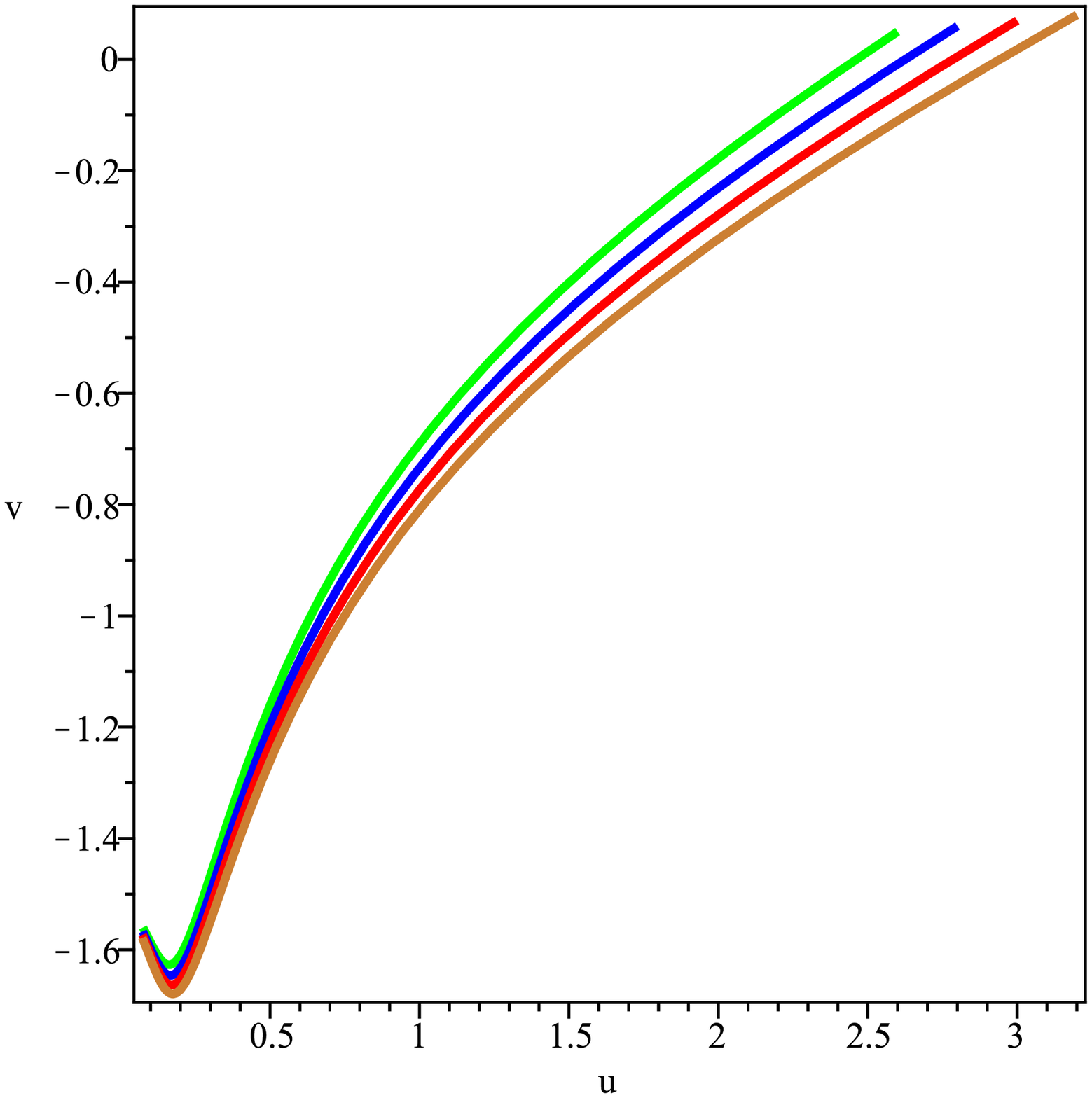}\\
\vspace{2mm}
~~~~~Fig.7~~~~~~~~~~~~~~~~~~~~~~~~~~~~~~~~~~~~~~~~~~~~~~~~~~~~~~~~~~~~~~~~~~~~~~Fig.8\\
\vspace{4mm}
\includegraphics[scale=0.35]{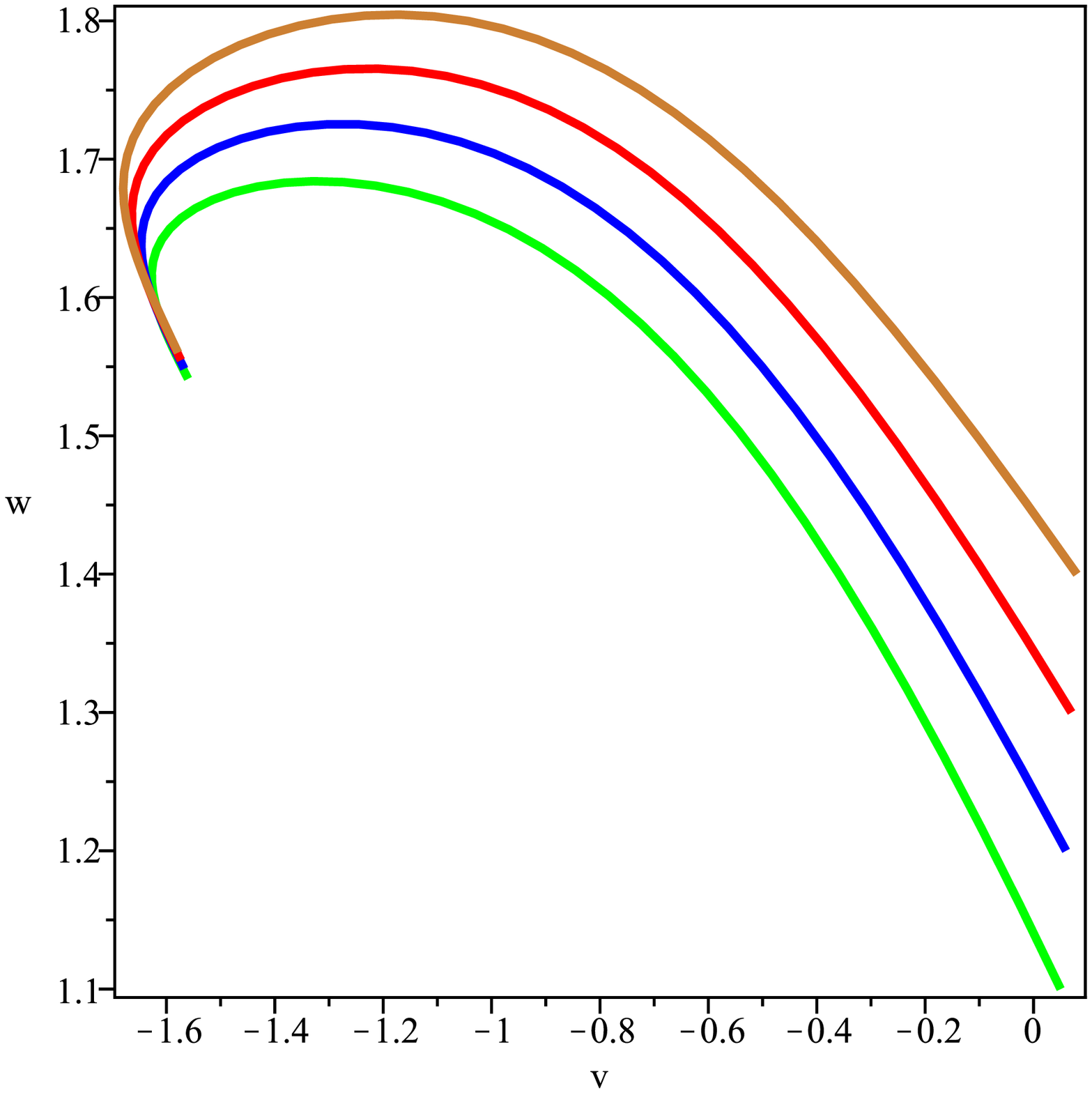}~~~~
\includegraphics[scale=0.35]{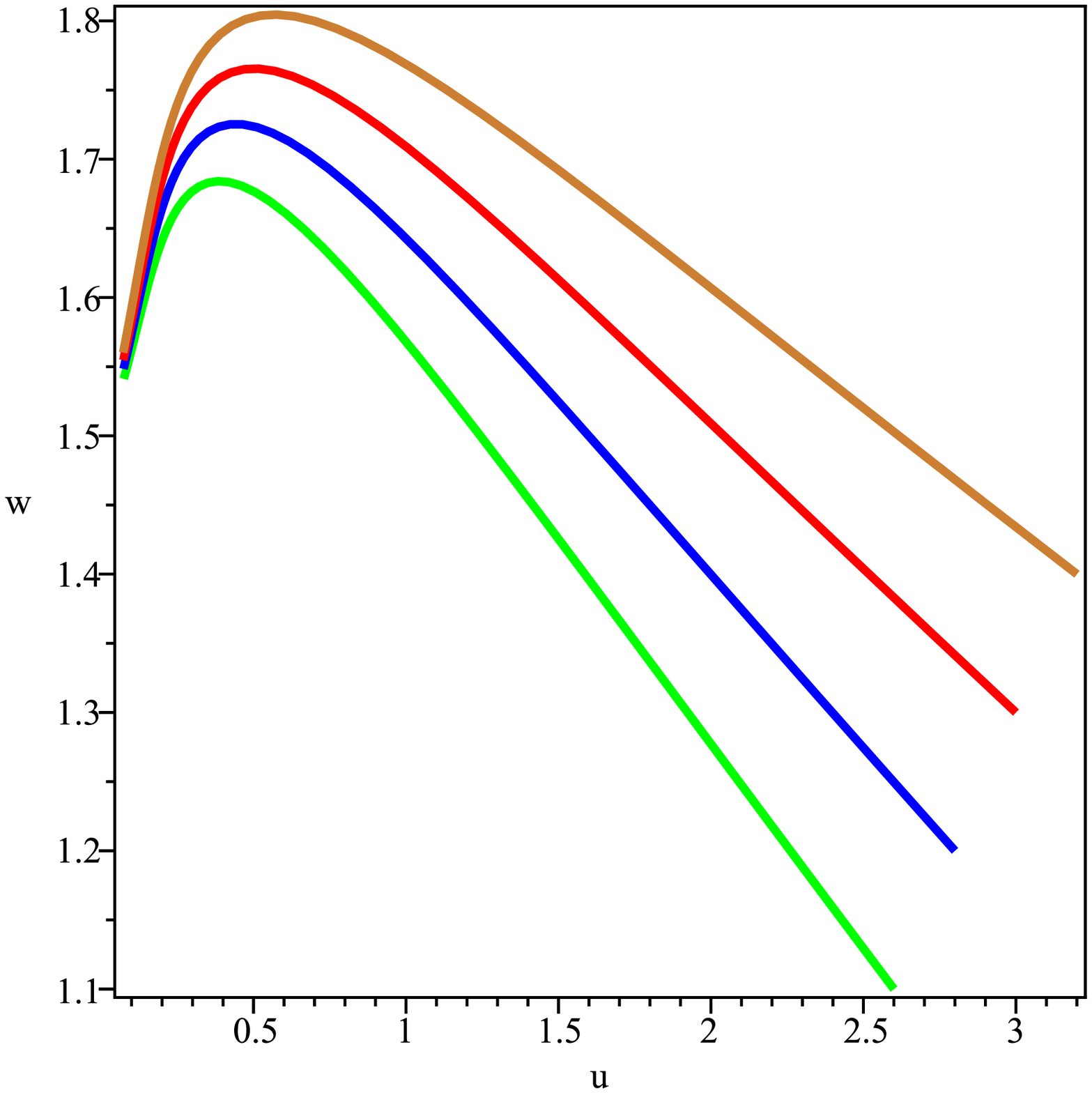}\\
\vspace{2mm}
~~~~~~~~Fig.9~~~~~~~~~~~~~~~~~~~~~~~~~~~~~~~~~~~~~~~~~~~~~~~~~~~~~~~~~~~~~~~Fig.10\\
\vspace{6mm} Fig. 7 shows the variations of the dimensionless
parameters $u$,$v$ and $w$ against the time for the initial
conditions $u(1.1)=2.5, v(1.1)=0.05,$ and $w(1.1)=2.8$. Fig.
8,9,10 show the phase space diagram in the sense of
$[v(t),u(t)]$,$[w(t),v(t)]$ and $[w(t),u(t)]$ respectively.
\vspace{6mm}
\end{figure}

\begin{figure}
\includegraphics[scale=0.45]{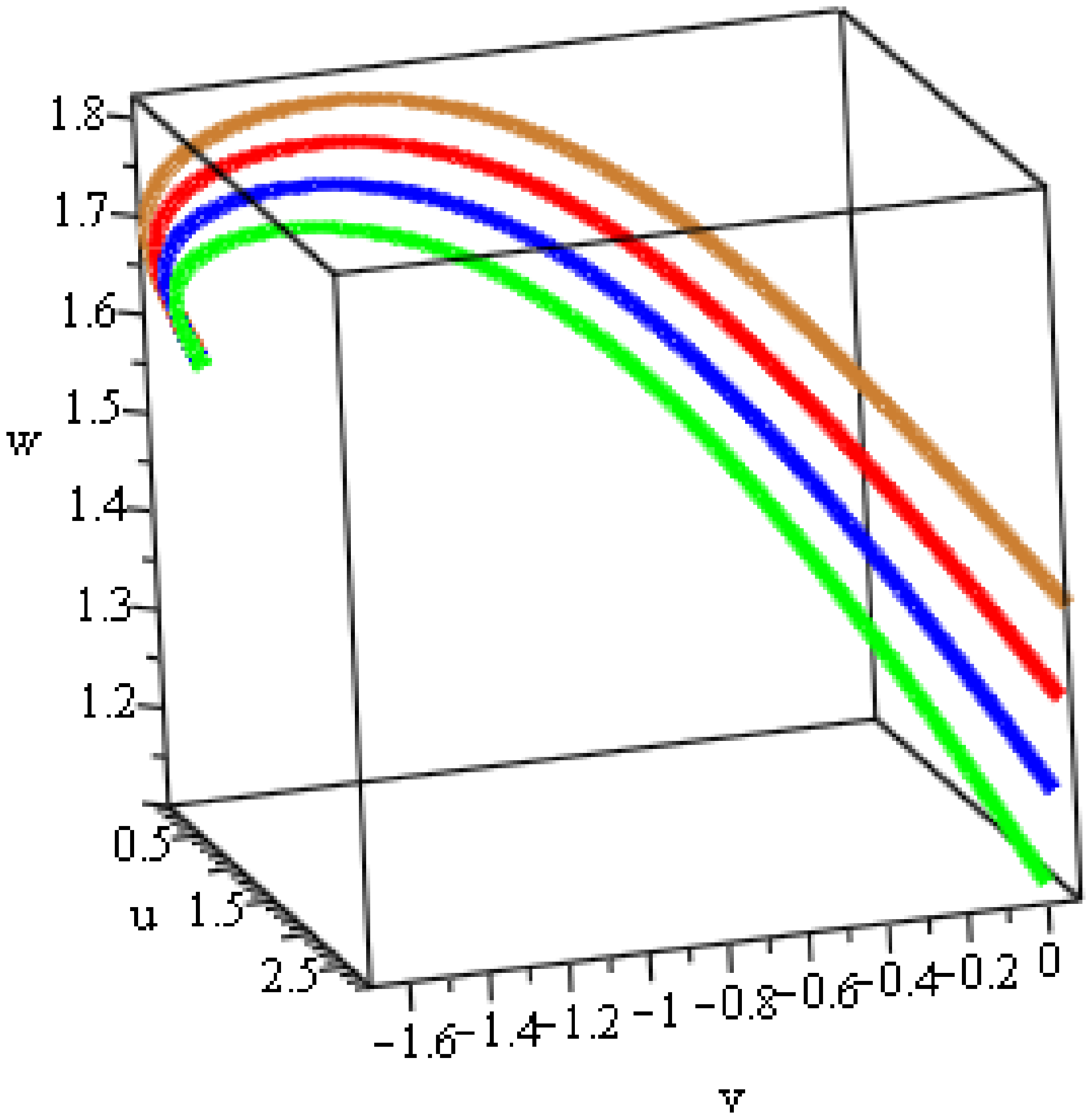}
\vspace{2mm}
~~~~~~~~~~~~~~~~~~~~~~~~~~~~~~~~~~~~~~Fig.11~~~~~~~~~~~~~~~~~~~~~~~~~~~~~~~~~~~~~~~~~~~~~~~~~\\
\vspace{4mm}
\includegraphics[scale=0.75]{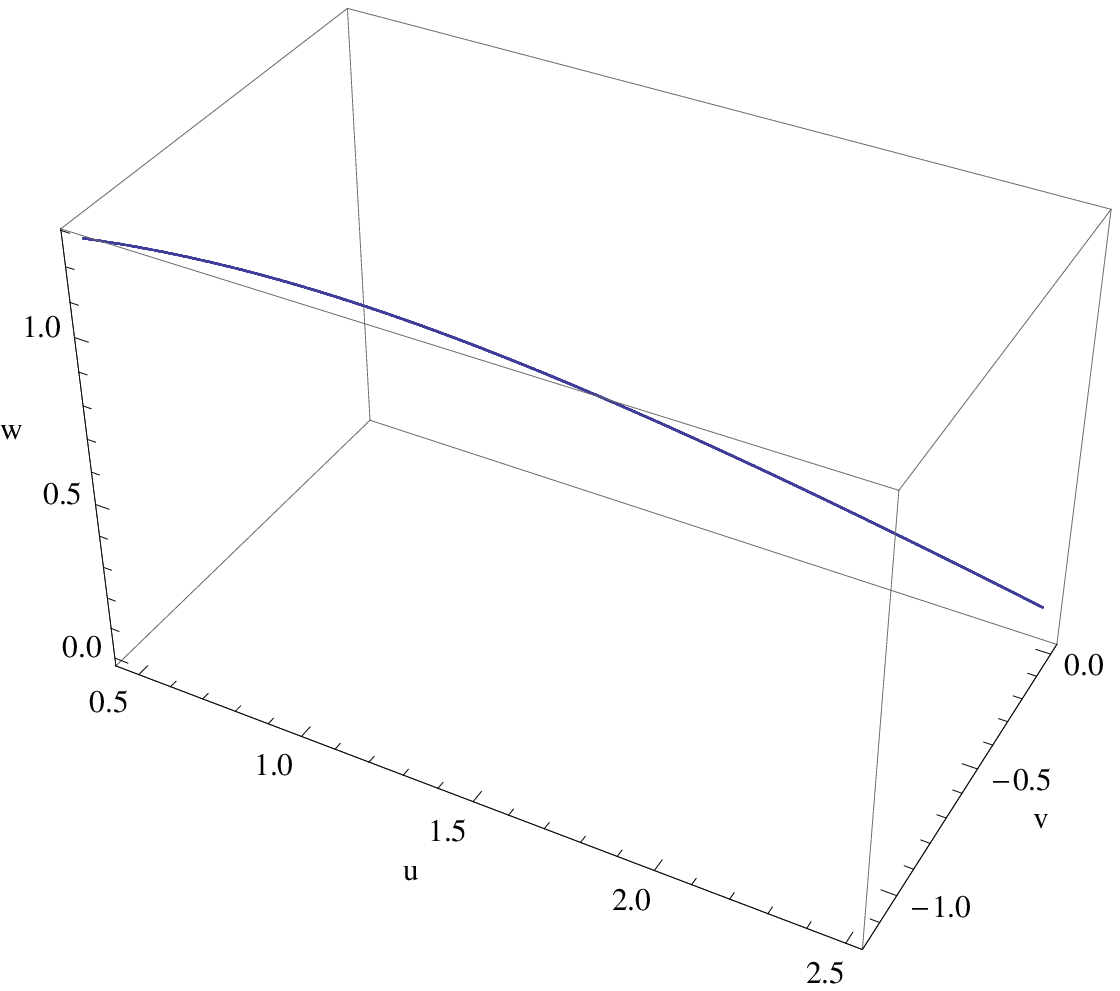}~~~
\includegraphics[scale=0.35]{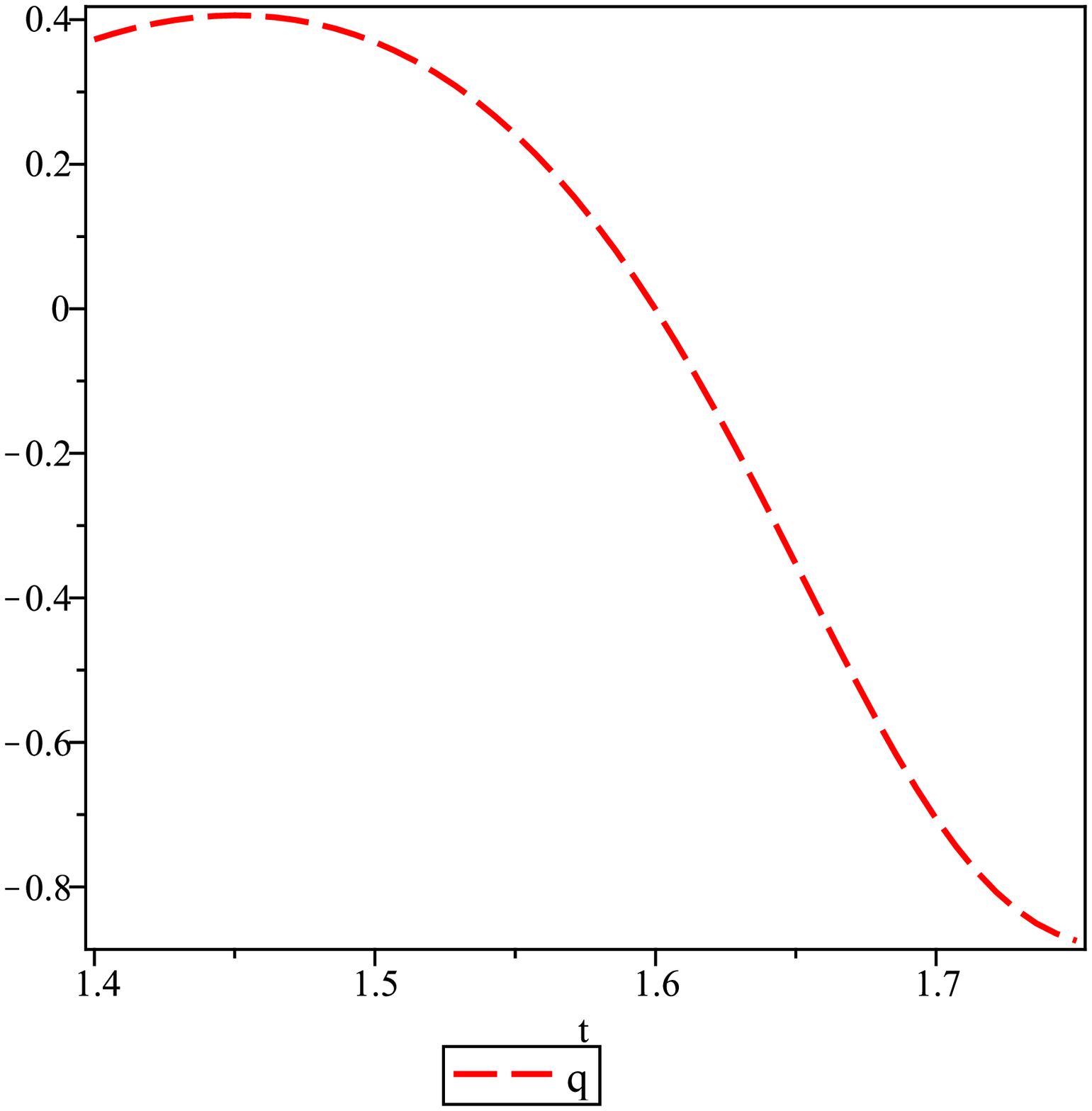}
\vspace{2mm}
~~Fig.12~~~~~~~~~~~~~~~~~~~~~~~~~~~~~~~~~~~~~~~~~~~~~~~~~~~~~~~~~~~~~~~~~~~~~~~~~~~~~~~Fig.13\\
\vspace{6mm} Fig. 11 and Fig. 12 show the 3D variations of the
dimensionless parameters $u$,$v$ and $w$ for four different
initial conditions and single initial condition. Fig. 13 shows the
variation of deceleration parameter $q$ against the time $t$.
\vspace{6mm}
\end{figure}

If the real parts of the above eigenvalues are negative at
critical points then critical points are stable and are stationary
attractor solutions; otherwise unstable and thus oscillatory.
Therefore the physical meaningful ranges are $n<\frac{8}{3} $ ; $
 c>0$ and $\frac{-3n^{2}+14n+4c-16}{-6n+12c+16}<\omega_{dm}\leq
\frac{1}{3}(3c+1)-\frac{\sqrt{9n^{2}-48n+36c^{2}+64}}{6}$ and in
this range the critical point $(u_{crit}, v_{crit}, w_{crit})$ is
stable and is a late-time stationary attractor solution.\\

The dimensionless parameters $u,v,w$ are drawn in figure 7 and we
see that these are decreasing in progression of time, whereas
$u,w$ keep positive orientation and $v$ keeps negative sign in
late time evolution. Also $v$ against $u$, $w$ against $v$ and $w$
against $u$ are drawn in figures 8, 9 and 10 respectively. Here
$v$ keeps negative sign when $u$ and $w$ increase. The 3D figure
of $u,v,w$ are drawn in figures 11 and 12 to get the visualization
of the natures of $u,v$ and $w$ for 4 different initial conditions
and a single initial condition respectively. The deceleration
parameter $q$ from figure 13 shows that $q$ decreases from $.5$ to
$-1$ which interprets the early deceleration and the late time
acceleration generated by the
VMCG in Kaluza-Klein Cosmology. \\

\section{\normalsize\bf{CONCLUSIONS}}

In this work, we have assumed the 5D Kaluza-Klein Cosmology for
anisotropic universe where the universe is filled with variable
modified Chaplygin gas (VMCG). The function $B$ is assumed as a
function of the scale factors $a$ and $b$ with some suitable
choice $B=B_{0}\left(a^{3}b\right)^{-n}$. If the energy density
and the pressure corresponding to the normal scalar field $\phi$
and the self interacting potential $V(\phi)$ can be associated
with the energy density and the pressure of the VMCG, the
corresponding $\phi$ and $V$ have been calculated in terms of the
scale factors $a$ and $b$. The variations of $V$ against $a$ and
$b$ are drawn in figures 1 and 2 and have seen that $V$ decreases
as universe expands. Also from figure 3, we see that $\phi$ always
increases as time goes on. Also we graphically analyzed the
geometrical parameters named {\it statefinder parameters} in
anisotropic Kaluza-Klein model. Fig. 4 and Fig. 5 show the
variations of $r$ against $a$ and $b$ and the variations of $s$
against $a$ and $b$ and Fig. 6 shows the variation of $r$ against
$s$ respectively for some particular choices of the constants.
From the figures, we conclude that $r$ increases and $s$ decreases
as $a$ and $b$ increase. From figure 6, we see that $s$ increases
from some positive value to $+\infty$ as $r$ decreases from finite
positive range. Also after certain stage, $s$ increases from
$-\infty$ to $0$ as $r$ decreases from
some positive to negative value.\\

Next, we consider a Kaluza-Klein model of interacting VMCG with
dark matter in the framework of Einstein gravity. Here we
construct the three dimensional autonomous dynamical system of
equations for this interacting model with the assumption that the
dark energy and the dark matter are interact between them and for
that we also choose the interaction term $Q$ in the form
$4bH\rho$, which is of course decreasing with the expansion of the
universe. We convert that interaction terms to its dimensionless
form and perform stability analysis and solve them numerically.
The critical point $(u_{crit},v_{crit},w_{crit})$ have been found
for the autonomous system. We obtain a stable scaling solution of
the equations in Kaluza-Klein model and graphically represent
solutions and have shown that the critical point is a late time
stable attractor. The dimensionless parameters $u,v,w$ are drawn
in figure 7 and we see that these are decreasing in progression of
time, whereas $u,w$ keep positive orientation and $v$ keeps
negative sign in late time evolution. Also $v$ against $u$, $w$
against $v$ and $w$ against $u$ are drawn in figures 8, 9 and 10
respectively. Here $v$ keeps negative sign when $u$ and $w$
increase. The 3D figure of $u,v,w$ are drawn in figures 11 and 12
to get the visualization of the natures of $u,v$ and $w$ for 4
different initial conditions and a single initial condition
respectively. The deceleration parameter $q$ from figure 13 shows
that $q$ decreases from $.5$ to $-1$ which interprets the early
deceleration and the late time
acceleration generated by the VMCG in Kaluza-Klein Cosmology. \\

\textbf{Acknowledgement:}\\

The authors are thankful to IUCAA, India for warm hospitality
where part of the work was carried out.\\

\end{document}